# Big Data Security Issues and Challenges in Healthcare

Behnam Kiani Kalejahi[1,2], Saeed Meshgini[1], Ayshan Yariyeva[2], Dawda Ndure[2], Uzeyir Maharramov[2], Ali Farzamnia[3]

[1] Department of Biomedical Engineering, Faculty of Electrical and Computer Engineering, University of Tabriz, Tabriz, Iran,
[2] Department of Engineering and Applied Sciences, Khazar University, Baku, Azerbaijan,
[3] Faculty of Engineering, University Malaysia Sabah, Kota Kinabalu, Sabah, Malaysia

*Abstract*—This paper embodies the usage of Big Data in Healthcare. It is significant to note that big data in terms of Architecture and implementation might be or has already or will continue to assist the continuous growth in the field of healthcare. The main important aspects of this study are the general importance of big data in healthcare, the positives big data will help tackle and enhance in this field and not to also forget to mention the tremendous downside big data has on healthcare that is still needed to improve or putting extensive research on. We believe there is still a long way in which institutions and individuals understand the hidden truth about big data. We have highlighted the various ways one could be confidently relied on big data and on the other hand highlighted the weighted importance of "big problem" big data and expected solutions.

## I. Introduction

The production capacity of computers is still growing, but not as rapidly as ten years ago compared to the amount of digital information that is growing at an alarming rate. In recent years, big data has become one of the dominant trends in the development of information technology. It is assumed that working with colossal amounts of unstructured data will have the greatest impact on production, government, trade, and medicine. An array of information is evaluated as big data only if it's so large that it becomes difficult to store, process and analyze. In fact, by the term Big Data, experts mean not a specific huge amount of data, but methods of processing them. The first biomedical revolution was associated with the emergence of microscopy and the introduction of a scientific approach to clinical research. Experts predict that big data will make a "second revolution" and will lead to a huge qualitative leap in the healthcare system. Without going into the wilds of definitions and schemes, we will analyze in a more detailed way the practical results of Big Data in healthcare. Why should we use big data in healthcare?

## II. ISSUES AND CONCERNS

The booming of healthcare [2] sector prompts the increased necessity to manage patient care and medical innovation. Machine learning algorithms, which can find statistical correlations in a huge global medical data file, will promptly issue predictions and recommendations for the patient and his attending Doctor or Physician. Basic strategies for applying big data in medicine create registries of medical data in which we can share information. Using this accumulated information, we can predict the possible "waves" of diseases.

### A. Prediction of disease development:

These electronic medical cards have already allowed doctors to establish a connection between seemingly fundamentally different diseases [3]. The risk assessment system developed in 2013 by the Kaiser Permanente consortium allows the medical practitioner to make a prediction about the development of dementia in patients with diabetes. By using the same model, the US military is trying to reduce the number of suicides among war veterans.

### B. Identification of genetic markers in oncology:

Scientists from the University of Cape Town UCT, analyzing the most common types of cancer, concluded that each of these types of cancer is characterized by a distinct combination of genes. It turns out that breast, intestinal, lung, ovarian and brain cancers are distinct genetic markers. According to the study leader, the team would not be able to make the discovery if it did not have access to an array of big data.

### C. Predicting the health of babies:

Toronto Children's Hospital has implemented Project Artemis. The hospital information system collects and analyzes data on infants in real-time. The system can track every 1260 indicators of the state of each child. This allows doctors to predict the unstable state of each child thereby allowing Doctors and Nurses to start the prevention of diseases in children.

### D. Prediction of risk factors in surgery:

Doctors at the Massachusetts General Hospital use QPID (qpid.apache.org. Apache Qpid is an open-source messaging system which implements the Advanced Message Queuing Protocol (AMQP)) analytical system to monitor important patient information throughout the course of their treatment. Another application of the QPID system in healthcare is the prediction of surgical risk [4].

The QPID system automatically searches for treatment protocols, after which it displays the results on a screen with a calculated red, yellow or green risk indicator.

*E. Creating new drugs:*

The greatest effect of Big Data is expected in the modeling of new drugs. Today with the help of data gathered from the medical field, Semantic Hub an IT company is assessing and developing the prospects for the manufacturing of new medical drugs.

*F. Improving the quality of clinical trials:*

Using Big Data technologies, companies can make clinical trials more effective. With the availability of patients' data from several databases, analytical systems can select patients who best meet the preliminary requirements of a drug test [6]. Thanks to the achievements of telemedicine, gadgets and wearable devices, researchers can monitor and track volunteers in real-time.

*G. Identification of drug side effects:*

Big data can also be used in predicting side effects for specific compounds and components even before the start of clinical trials. By using an analytical method companies can save time, money and save patients' lives by checking dozens of different drug characteristics before putting it out in the market. Data on the actual practice of drug use is collected outside the framework of traditional randomized clinical trials, which today are the basis of drug testing, and interest in this area is growing rapidly. The number of tests with the analysis of big data for the year 2017 exceeded 300, reports Reuters (referring to the international clinical research website clinicaltrials.gov). Why does Big Data appear to be a big problem?
It would be wrong to argue that big data is the key to absolutely all useful knowledge. When extracting the necessary information from a vast array of data, several significant difficulties arise.

*H. Unstructured data:*

For textual information, a well-developed search algorithm can be used to extract needed information. The problem arises when the information is in the form of a speech or a video. The technology of turning a speech or a video into a text can be realized, but the data volumes will be extremely large thereby causing storage and sorting issues. Approximately 78% of medical data is not structured, filtering and analyzing such amounts of information will be high-costly.

*I. High risk of information distortion:*

Some critics even believe that Big Data is one big hoax. A flurry of indignation hit big data after the acclaimed failure of Google Flu Trends. For example, the project from Google missed the 2013 epidemic in the US and distorted information about it by 140%. Scientists from several American universities then found that in the past two years of work, the analysis more often showed incorrect results. Frameworks related to big data in healthcare systems.

**IBM Watson**: Petabytes of medical information form arrays of big data. All this information can be "fed" to the IMB Watson supercomputer. After analyzing the data taken from the treatment of many patients, the supercomputer helps the doctor to choose the best treatment for an individual. In the spring of 2015, Apple and IBM announced a joint project on the use of big data in healthcare. The two companies are working on a single platform that allows owners of the iPhone and Apple Watch to collect and send personal health information to Watson Health - an IBM medical analytics service.

The data size of genomics is 2-40 EB/year which can be stored in local databases or in the cloud. Cloud computing is used for storage, distribution, and processing of the data, where it can be utilized, so that applications can run on remote machines that already have access to data. A data platform that integrates genomics/healthcare data while enabling quick and efficient analysis would allow extraction of practical insights in a short frame of time.

Apart from the frameworks, there are many programming languages and graphic models as well that contribute to the usage of Big Data in Healthcare: SQL, NoSQL, MapReduce, Girap, GraphLab, Hadoop [7], etc. However, coming up with an in-depth approach, the SQL is not an appropriate language for scalable analytics whilst NoSQL is not optimized for reading data. MapReduce is scalable, but it does not support iterative analytics. Some large-scale graph processing systems such as Giraph and GraphLab provide support for data consistency by providing configurable consistency models.

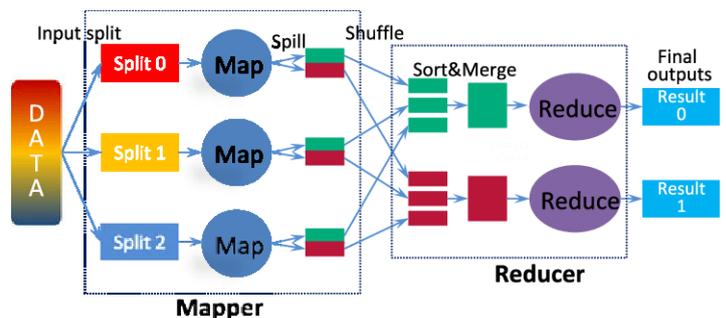

Figure. 1 Hadoop MapReduce framework [8]

As you can see from the image (Fig 1) with input data entering into the system it is then split up and mapped into different data packages, and finally, it is sorted, merged and reduced which will have the final output evaluated as 0's and 1's.

### III. METHODOLOGY

The transition to modern information technologies in healthcare helps improve the quality of service, shortens the examination time, increases the accuracy of diagnostics, makes it possible to conduct remote consultations, examinations, remotely process primary information, able to store patient data in digital form for a long time and, if necessary, access them from anywhere in the world [9]. Currently, the main directions of e-health development are advisory networks for health workers and patients, electronic medical records (medical records) system, medical insurance, pharmacy information, and ambulance dispatch systems. Recently, information systems related to telemedicine have acquired relevance [10]. The process of telemedicine counseling is accompanied by the transfer of applications, extracts, medical images, conclusions, legal and financial documents between the participants. The intellectual basis of telemedicine complexes is medical information systems (MIS) [11]. They integrate and store all the data allowing you to search and analyze the necessary information to the doctor.

The development of modern medicinal systems is directly related to the presence of high-speed communication channels in this field, which allows using, if necessary, the methods of distributed data processing ODP (Online Distributed Processing) [12]. Communication problems in some regions can be significant and, in some cases, these are not only "last mile" problems but also problems connecting users to regional and global networks. The prevalence of full-featured systems is currently insufficient. The current practice is aimed at automating fiscal and partly reporting functions rather than reducing information ambiguity in order to increase the efficiency of medical decisions and implement procedures in accordance with the standards of treatment and the efficiency of using health care facilities [13]. Individual fragments of medical information are processed by different personalities of various health facilities - the registrar, doctor, diagnostician, nurse, statistician, laboratory technician, etc. Moreover, many of these fragments can be processed separately thereby representing information which is not uniquely associated with an individual.

A special feature of medical information is its confidentiality. A number of data entered, processed and stored in the operation of medical information systems are personal data or may constitute medical confidentiality. In addition, the database of the medical information system contains critical information on which, often, a person's life can depend, therefore ensuring database integrity is of paramount importance as well as the ability to monitor the state of the system itself and its security. However, increasing security can negatively affect the speed and reliability of the software, therefore, the concept of information security MIS should be based on the specifics of the health facilities as a queuing system.

## IV. SUGESTED SOLUTIONS

It is important to note that the complex computerization of medical institutions, the development of specialized integrated IT systems, however, requires the study of a large number of specific issues such as problems of electronic document management, problems of standardization of information representation, problems of choice and/or development of software and database management systems; issues of intellectualization of the database, the formation of "operational" and "analytical" forms of information; problems of reliability, security, confidentiality, privacy; problems of transition to digital technologies; mobility issues; problems of "origin" of information etc.

Secondly, the material for analysis and subsequently the result of the analysis can be tied by the laboratory technician to the number of the medical card (bar code, unique key, etc.) and at the same time the identification of the person to whom the medical card belongs does not occur due to the fact that health information systems of different regions are designed and developed by a non-unified methodology hence, therefore, reviewing and analyzing the activities of the health system as a whole becomes impractical because the protection system (building information security) is decentralized. In this regard, a transition is required from fragmented systems to complex medical information systems at the level of the Central Medical Institution (health governing body) including insurance medical companies which solves all the tasks of its information interactions.

Moreover, to build a system of information protection in medical systems, it is necessary to consider each of the stages of its processing: from the moment data arrives at the system, during initial circulation, until destruction, when the storage period expires. Information enters the system when the patient first turns to a medical institution — at this stage, the patient's data, the results of the primary diagnostics, the prescribed medications and procedures are collected and processed.

The second stage is the processing of information, the transfer of data into a standard electronic format and entry into a database and the last step is to store the information in a database while the medical system may periodically access the stored information. When a patient reverts to this medical institution, information about him/her in the database is updated and adjusted and data for medical statistics are collected. After the expiration of the established storage period, the information in the system is utilized or transferred to distant archives - consolidated.

On a final note, as a necessary measure to protect against external attacks, it is advisable to introduce control over servers, switches, and workstations for unusually high activity, full use of anti-virus protection on servers and workstations as well as firewall implementations, and monitor all updates for existing operating systems [15]. Already at the stage of developing an architectural solution, it is necessary to introduce a classification of stored data according to their degree of importance and confidentiality, as well as using multi-level user authentication, involving the use of USB keys, smart cards, passwords, file keys, etc. In cases when the network architecture provides for remote access to data, firewalls should be used, IDS (Intrusion Detection System) systems and VPN should be in place. As an additional measure of protection, it is necessary to provide traffic monitoring devices (security scanners) in the network segment in order to detect DoS attacks. During the development phase of the SAN (Storage Area Network) architecture, it is important to maximize the use of device spacing between isolated areas using hardware zoning (Hard Zoning), and software zoning (Soft Zoning) or LUN masking all useful as additional protection. The mandatory rule should be the organization of a dedicated network segment to monitor and control devices in the SAN (Storage Area Network) network. The level of access control is the most likely target for various attacks in order to gain administrative access to NAS devices (Network Attached Storage - network storage), therefore one of the preventive security measures is the use of secure access protocols (in particular, SSH-Secure Shell or HTTPS). Integrated security of data on an automated workplace provides for the protection of temporary files while working with confidential information, especially during remote teleconsultations, sending information by E-mail, etc. Such security measures may include blocking TEMP directories, paging software applications, deleting temporary Internet files [16]. Proposed solutions for the organization of information security, the implementation of an IIA (Institute of Internal Auditors) makes it possible to ensure an acceptable level of information protection, without significant restrictions for the users of an IIA in their actions.

## V. CONCLUSION

Despite criticism, the introduction of big data into medical practice in Western countries is accelerating. The main technological promoter of this process is the ubiquitous transition to electronic medical records.

According to a HITECH (The Health Information Technology for Economic and Clinical Health Act) study, over 94% of hospitals in the US use electronic medical cards. Countries in Europe are a little bit lacking behind in this area, but the European Commission has issued a directive which is designed to fundamentally change the situation.

It is assumed that by 2020 the European centralized system of medical records will become a reality.

Deloitte Center consultants believe that by 2020 big data will completely change the healthcare field; thanks to gadgets, patients will know almost everything about their health, and will be able to personally participate in choosing the best treatment.

With the help of big data and machine learning, a learning health system will be developed, prediction will get easier, for example, the response of a patient to radiation therapy will be clarified unchallenged than it used to be.

## REFERENCES

[1] Y. Wen, J. Liu, W. Dou, X. Xu, B. Cao, J. Chen, Scheduling workflows with privacy protection constraints for big data applications on cloud, Future Generation Computer Systems doi: 10.1016/j.future.2018.03.028.

[2] Rev. Latino-Am. Enfermagem;25: e2848 DOI: 10.1590/1518-8345.1502.2848, 2017.

[3] Patterson, Karlyn & A. Lambon Ralph, Matthew. The Hub-and-Spoke Hypothesis of Semantic Memory. 10.1016/B978-0-12-407794-2.00061-4, 2016.

[4] Clinical and non-clinical compliance EMA/84016/2014, 3 July 2014.

[5] Patterson, Karlyn & A. Lambon Ralph, Matthew. The Hub-and-Spoke Hypothesis of Semantic Memory. 10.1016/B978-0-12-407794-2.00061-4, 2016.

[6] K. Yang, Q. Han, H. Li, K. Zheng, Z. Su, X. Shen, An efficient and fine-grained big data access control scheme with privacy-preserving policy, IEEE Internet of Things Journal 4 (2) (2017) 563–571.

[7] J. Li, D. Lin, A. C. Squicciarini, J. Li, C. Jia, Towards privacy-preserving storage and retrieval in multiple clouds, IEEE Transactions on Cloud Computing 5 (3) (2017) 499–509. doi:10.1109/TCC.2015.2485214.

[8] Park, Dongchul & Wang, Jianguo & Kee, Yang-Suk. In-Storage Computing for Hadoop MapReduce Framework: Challenges and Possibilities. IEEE Transactions on Computers. PP. 1-1. 10.1109/TC.2016.2595566, 2016.

[9] A. Razaque, S. S. Rizvi, Privacy preserving model: a newscheme for auditing cloud stakeholders, Journal of Cloud Computing 6 (1) (2017) 7.

[10] Y. Wang, X. Wu, D. Hu, using randomized response for differential privacy preserving data collection., in: EDBT/ICDT Workshops, Vol. 1558, 2016.

[11] Alshugran T, Dichter J. Extracting and modeling the privacy requirements from HIPAA for healthcare applications. IEEE Long Isl. Syst. Appl. Technol. LISAT Conf. 2014, p. 1–5. doi:10.1109/LISAT.2014.6845198, 2014.

[12] A. Alhayajneh, A. Baccarini, G. Weiss, T. Hayajneh, A. Farajidavar, Biometric authentication and verification for medical cyber physical systems, Electronics 7 (12) (2018) 436.

[13] I. H. Witten, E. Frank, M. A. Hall, C. J. Pal, Data Mining: Practical machine learning tools and techniques, Morgan Kaufmann, 2016.

[14] A. H. Sodhro, A. K. Sangaiah, S. Pirphulal, A. Sekhari, and Y. Ouzrout, "Green media-aware medical IoT system,"

Multimedia Tools and Applications, pp. 1-20, 2018.

[15] Z. Ali, M. Imran, M. Alsulaiman, M. Shoaib, S. Ullah, Chaos-based robust method of zero-watermarking for medical signals, Future Generation Computer Systems, Vol. 88 (11), pp. 400-412, Nov 2018.

[16] W. Sun, Z. Cai, Y. Li, F. Liu, S. Fang, and G. Wang, "Security and Privacy in the Medical Internet of Things: A Review," Security and Communication Networks, vol. 2018.

[17] Z. Ali, M. Imran, M. Alsulaiman, T. Zia, M. Shoaib, A zero-watermarking algorithm for privacy protection in biomedical signals, Future Generation Computer Systems, Vol. 82, pp. 290-303, May 2018.